\magnification \magstep1
\raggedbottom
\openup 2\jot
\voffset6truemm
\leftline {\bf A SPECTRAL APPROACH TO YANG-MILLS THEORY}
\vskip 1cm
\leftline {Giampiero Esposito}
\vskip 1cm
\noindent
{\it Istituto Nazionale di Fisica Nucleare, Sezione di Napoli,
Complesso Universitario di Monte S. Angelo, Via Cintia,
Edificio N', 80126 Napoli, Italy}
\vskip 0.3cm
\noindent
{\it Dipartimento di Scienze Fisiche, Universit\`a di Napoli
Federico II, Complesso Universitario di Monte S. Angelo, Via
Cintia, Edificio N', 80126 Napoli, Italy}
\vskip 1cm
\noindent
{\bf Abstract}. Yang--Mills theory in four dimensions is studied by using the
Coulomb gauge. The Coulomb gauge Hamiltonian involves integration
of matrix elements of an operator $\cal P$ built from the Laplacian
and from a first-order differential operator. The operator $\cal P$ is
studied from the point of view of spectral theory of pseudo-differential
operators on compact Riemannian manifolds, both when self-adjointness holds
and when it is not fulfilled. In both cases, well-defined matrix elements
of $\cal P$ are evaluated as a first step towards the more difficult 
problems of quantized Yang--Mills theory.
\vskip 100cm
\leftline {\bf 1. Introduction}
\vskip 0.3cm
\noindent 
Although much progress has been made in the theoretical understanding
of gauge theories of fundamental interactions,$^{1-3}$ a number of
outstanding unsolved problems remain. In particular, the 
{\it mass gap problem} has been stressed in the recent literature,$^{4}$ 
which amounts to proving that, for any compact simple gauge group,
quantum Yang--Mills theory on ${\bf R}^{4}$ has an Hamiltonian operator
with no spectrum in the interval $(0,\delta)$ for some $\delta >0$.
One should then show that, starting from the classical
action functional
$$
I=-{1\over 4}\int {\rm Tr}(F_{\mu \nu}F^{\mu \nu})d^{4}x,
\eqno (1)
$$
the corresponding Hamiltonian operator has energy spectrum bounded
from below, with strictly positive lower bound. The solution of such
a problem is very important not only to achieve internal consistency
of the mathematical formalism, but also (if not mainly) for physical
reasons: the existence of a mass gap would account for the nuclear
force being strong but short-ranged. Moreover, the recent theory of
glueballs$^{5}$ relies entirely on action functionals like (1),
without fermionic fields. In plain terms, it is possible for two or
more gluons to combine into a strongly bound, neutral-coloured
particle of pure glue. This (hypothetical) object is called a glueball.$^{5}$
Moreover, a gluon can bind with a meson to form a {\it hybrid}. For example,
a red quark and an anti-blue antiquark can bind with a blue/antired
gluon. The lightest glueball allowed by quantum chromodynamics can
be described by a circular tube of glue and has vanishing angular
momentum.$^{5}$ This has spherical symmetry while glueballs of other,
elongated shapes have non-vanishing angular momenta and larger masses.  
Hybrids can announce their presence by yielding at least three 
$s$-wave mesons. They are in fact predicted to decay into one 
$s$-wave meson and another short-lived meson with internal angular
momentum. The latter then breaks up into two $s$-wave mesons. 
The first (though uncertain) 
experimental evidence in favour of hybrids was obtained in
1994, when experimenters at Protvino found an object called 
$\pi(1800)$, emerging from collisions of pions with protons. This
particle has the quantum properties and decay pattern expected 
for a hybrid. Recent research deals with special hybrids called
exotics, which have combinations of internal angular momentum,
parity and charge conjugation quantum numbers that are forbidden
for mesons. For example, the simplest exotic has $J=1,P=-1$ and $C=1$.
Inconclusive evidence also exists in favour of glueballs called
$f_{0}(1500)$ and $f_{j}(1710)$, which would belong to the class of
glueballs of mass in the range 1.500 to 1.800 MeV.$^{5}$ 
The latest experiments tell us that glueball candidates and 
$q{\overline q}$ mesons have been found to be produced with different
momentum and angular dependences in the central region of
$pp$ collisions.$^{6}$

All this phenomenology can be described, in principle, with
the help of a Yang--Mills Lagrangian. We have been therefore motivated,
in our research, by the mass gap problem in four dimensions, although
it will become clear that, for the time being, we only have a possibly
new perspective in classical Yang--Mills theory.
For this purpose, we have considered the Yang--Mills
Lagrangian in the Coulomb gauge with the associated decomposition
into electric and magnetic parts. Although the Coulomb gauge is
non-covariant, it has the merit of leading to a quantum theory which
is manifestly unitary. Moreover, well-established calculational 
recipes are available for writing down the Hamiltonian operator of
the quantum theory,$^{7}$ while recent work has shown that the Coulomb
gauge can be viewed as the singular limit of the Landau--Coulomb 
interpolating gauge,$^{8}$ adding therefore evidence in favour of such a
gauge being very appropriate for the quantum theory.

Section 2 presents a review of the Coulomb gauge Hamiltonian, with 
emphasis on matrix elements of an operator $\cal P$ which plays a key 
role in Sec. 3. Here, attention is focused on compact 
Euclidean spacetime, for which discrete spectral resolutions of self-adjoint
elliptic operators exist. The matrix elements of $\cal P$ among square-integrable
functions are evaluated explicitly in such a case. Section 4 presents an
assessment of the spectral approach to the mass gap.
\vskip 0.3cm
\leftline {\bf 2. Coulomb Gauge Hamiltonian}
\vskip 0.3cm
\noindent
To help the general reader, we present here a brief review of the
Coulomb gauge Hamiltonian in classical and quantum theory, relying 
on Ref. 7.

The Lagrangian $L$ of a classical Yang--Mills field can be expressed
in terms of electric and magnetic fields, which are defined by
$$
{\vec {\cal E}}_{i}  \equiv -{d\over dt}{\vec A}_{i}
-(\nabla_{i}+g{\vec A}_{i} \times){\vec A}_{0} 
=-{d\over dt}{\vec A}_{i}+{1\over g}{\cal D}_{i}
{\vec \omega},
\eqno (2)
$$
$$
\varepsilon_{jki} {\vec {\cal B}}_{i} \equiv \nabla_{j}{\vec A}_{k}
-\nabla_{k}{\vec A}_{j}+g{\vec A}_{j} \times {\vec A}_{k}.
\eqno (3)
$$
In Eq. (2), ${\vec A}_{0}=-{{\vec \omega}\over g}$, and
${\cal D}_{i}$ is the covariant derivative in the Coulomb gauge:
$$
{\cal D}_{i} \equiv \nabla_{i}+g {\vec A}_{i} \times,
\eqno (4)
$$
the cross denoting the isovector product in all our equations.
Thus, the Lagrangian reads
$$
L={1\over 2}\int \Bigr({\vec {\cal E}}_{i} \cdot {\vec {\cal E}}_{i}
-{\vec {\cal B}}_{i} \cdot {\vec {\cal B}}_{i}\Bigr)d^{3}r,
\eqno (5)
$$
with corresponding Hamiltonian
$$
H={1\over 2}\int \Bigr({\vec {\cal E}}_{i} \cdot {\vec {\cal E}}_{i}
+{\vec {\cal B}}_{i} \cdot {\vec {\cal B}}_{i}\Bigr)d^{3}r.
\eqno (6)
$$
One can now exploit the constraint $\nabla_{j}{\vec A}_{j}=0$ and
decompose the electric field into a transverse part and a gradient
term, i.e.
$$
{\vec {\cal E}}_{i}={\vec {\cal E}}_{i}^{\rm tr}
-\nabla_{i}{\vec \phi},
\eqno (7)
$$
where (hereafter we define
the Laplacian as the operator $\bigtriangleup \equiv -\nabla^{k}\nabla_{k}$,
with a minus sign in front of second derivatives to make it 
bounded from below)
$$
{\vec {\cal E}}_{i}^{\rm tr}=-{d\over dt}{\vec A}_{i}
+\Bigr(\delta_{ij}+\bigtriangleup^{-1} \nabla_{i}\nabla_{j}\Bigr)
({\vec A}_{j} \times \omega),
\eqno (8)
$$
which satisfies $\nabla_{i}{\vec {\cal E}}_{i}^{\rm tr}=0$, and
$$
{\vec \phi}={\vec A}_{0}-g \bigtriangleup^{-1}\Bigr({\vec A}_{j}
\times \nabla_{j}{\vec A}_{0}\Bigr).
\eqno (9)
$$
Since the electric field has vanishing divergence with respect to the
connection $\cal D$, i.e. ${\cal D}_{i}{\vec {\cal E}}_{i}=0$, one
gets also an equation for $\vec \phi$, i.e.
$$
-{\vec A}_{i} \times {\vec {\cal E}}_{i}^{\rm tr}
+{1\over g}\nabla_{i}{\cal D}_{i}{\vec \phi}=0.
\eqno (10)
$$
Such an equation is solved upon inverting the operator 
$\nabla_{i}{\cal D}_{i}$, which yields (summation over
repeated indices is understood)
$$
{\vec \phi}=g(\nabla_{k}{\cal D}_{k})^{-1}
{\vec A}_{i} \times {\vec {\cal E}}_{i}^{\rm tr}
=-g(\nabla_{k}{\cal D}_{k})^{-1}
{\vec A}_{i} \times {\vec \Pi}_{i}^{\rm tr}.
\eqno (11)
$$
At this stage our analysis is still classical, but in the quantum theory
the counterpart of Eq. (11) is more involved because zero-modes of the
Faddeev--Popov operator play a role and their effect should be included.

Upon exploiting the transverse nature of ${\cal E}_{i}^{\rm tr}$
and imposing fall-off conditions at infinity on $\vec \phi$,
the classical Hamiltonian reads
$$
H={1\over 2}\int \left[\Bigr({\vec {\cal E}}_{i}^{\rm tr}\Bigr)^{2}
+\Bigr(\nabla_{i}{\vec \phi}\Bigr)^{2}
+\Bigr({\vec {\cal B}}_{i}\Bigr)^{2}\right]d^{3}r.
\eqno (12)
$$
This is further simplified by using the identity 
$$
(\nabla_{i}\vec \phi)(\nabla_{i}\vec \phi)
=\nabla_{i}({\vec \phi}\nabla_{i}{\vec \phi})
+{\vec \phi} \bigtriangleup {\vec \phi},
\eqno (13)
$$
jointly with Eq. (11). Thus, on defining
$$
{\vec \Pi}_{i}^{\rm tr},
\eqno (14)
$$
which represents the charge carried by ${\vec A}_{i}$, the classical
Hamiltonian reads eventually
$$ \eqalignno{
H&= {1\over 2}\int \left[\Bigr({\vec \Pi}_{i}^{\rm tr}\Bigr)^{2}
+\Bigr({\vec {\cal B}}_{i}\Bigr)^{2}\right]d^{3}r \cr 
&+ {g^{2}\over 2} \int \sigma_{A}^{l}({\vec r}) \langle l,r |
{\cal P} | l',r' \rangle \sigma_{A}^{l'}({\vec r}')d^{3}r d^{3}r'.
&(15)\cr}
$$
In Eq. (15), $\cal P$ is the integro-differential operator defined by
$$
{\cal P} \equiv (\nabla_{i}{\cal D}_{i})^{-1} \bigtriangleup
(\nabla_{j}{\cal D}_{j})^{-1},
\eqno (16)
$$
where the inverse of $\nabla_{i}{\cal D}_{i}$ results from (11), and the
Laplacian results from (13). The scheme obtained from (15) is not
amenable to calculation unless one finds a convenient way of expressing 
the inverse of $\nabla_{i}{\cal D}_{i}$. For this purpose, one defines
$$
\Lambda \equiv \nabla_{i}({\vec A}_{i} \times )
={\vec A}_{i} \times \nabla_{i},
\eqno (17)
$$
so that
$$
\nabla_{i}{\cal D}_{i}=-\bigtriangleup +g \Lambda,
\eqno (18)
$$
and hence 
$$ \eqalignno{
\; & (\nabla_{i}{\cal D}_{i})^{-1}= (-\bigtriangleup+g \Lambda)^{-1}
=-\bigtriangleup^{-1} \Bigr(I+g \Lambda \bigtriangleup^{-1} 
+(g \Lambda \bigtriangleup^{-1})^{2}+{\rm O}(g^{3})\Bigr) \cr 
&= -\bigtriangleup^{-1}-g \bigtriangleup^{-1}\Lambda \bigtriangleup^{-1}
-g^{2}\bigtriangleup^{-1}\Lambda \bigtriangleup^{-1} \Lambda \bigtriangleup^{-1}
+{\rm O}(g^{3}),
&(19)\cr}
$$
where we have applied the formula 
$$
(AB)^{-1}=B^{-1}A^{-1}
$$ 
to the operators $A \equiv I-g \Lambda \bigtriangleup^{-1}$ and 
$B \equiv -\bigtriangleup$.
The insertion of the (formal) expansion (19) into the definition (16)
yields an algorithm for $\cal P$, i.e.
$$
{\cal P}= \bigtriangleup^{-1}+\sum_{k=1}^{\infty}
(k+1)g^{k} \bigtriangleup^{-1}
{\left(\Lambda \bigtriangleup^{-1} \right)}^{k},
\eqno (20)
$$
which is useful at small $g$ and for the analysis of spectral asymptotics.
 
In the quantum theory, one has instead to consider the un-renormalized
coupling constant $g_{0}$ and the Faddeev--Popov determinant$^{7}$ 
$$
\gamma \equiv {\rm det} (\nabla_{i}{\cal D}_{i}).
\eqno (21)
$$
The equal-time canonical commutation relations read (the smeared form
is more rigorous but inessential for our purposes, which are
not axiomatic)
$$
\Bigr[A_{j}({\vec r},t)^{l},\Pi_{k}^{\rm tr}({\vec r}',t)^{m}\Bigr]
=i \delta^{lm}\Bigr(\delta_{jk}+\bigtriangleup^{-1}\nabla_{j}\nabla_{k}\Bigr)
\delta^{3}({\vec r}-{\vec r}'),
\eqno (22)
$$
$$
\Bigr[A_{i}({\vec r},t)^{l},A_{j}({\vec r}',t)^{m}\Bigr]
=\Bigr[\Pi_{i}^{\rm tr}({\vec r},t)^{l},
\Pi_{j}^{\rm tr}({\vec r}',t)^{m}\Bigr]=0,
\eqno (23)
$$
and the Hamiltonian operator in the Coulomb gauge takes the form$^{7}$
$$ \eqalignno{
{\hat H}&= \int \left[{1\over 2}\gamma^{-1}{\vec \Pi}_{i}^{\rm tr}
\cdot \gamma {\vec \Pi}_{i}^{\rm tr}
+{1\over 2}{\vec {\cal B}}_{i}^{2}\right]d^{3}r \cr 
&+ {g_{0}^{2}\over 2}\int \gamma^{-1}\sigma^{l}({\vec r})
\langle l,{\vec r}| {\cal P}| l',{\vec r}' \rangle
\gamma \sigma^{l'}({\vec r}')d^{3}r d^{3}r'.
&(24)\cr}
$$
Note that, since quantum fields are operator-valued distributions,$^{9}$ the
Hamiltonian as in the form just written is not defined as an operator
in a Hilbert space. Strictly, the product of local operators should be
regularized, and all quantum formulae should be written with this
understanding.
\vskip 0.3cm
\leftline {\bf 3. Structure of $\cal P$}
\vskip 0.3cm
\noindent
Our ultimate goal is the investigation of the spectrum of $\hat H$, 
with the associated role played by $\gamma,
{\vec \Pi}_{i}^{\rm tr},{\vec {\cal B}}_{i},\sigma^{l}$. But this is
still extremely difficult, and hence we resort to the analysis of the
operator $\cal P$ (see Eq. (16)) occurring in the classical theory,
here written in the form
$$
{\cal P}=(\bigtriangleup - g \Lambda)^{-1} \bigtriangleup
(\bigtriangleup -g \Lambda)^{-1},
\eqno (25)
$$
where one should bear in mind that $\Lambda$ is the first-order differential
operator defined in Eq. (17). Now two main cases can be distinguished in
a framework where Minkowski space-time is replaced by a compact 
four-geometry $(M,g)$ without boundary. This is more relevant for Euclidean
field theory, which is nevertheless an important branch of quantum field
theory.$^{10,11}$ By doing so, one may hopefully learn lessons about operators
whose analysis is a mandatory step before being able to solve 
the original problem.
\vskip 0.3cm
\noindent
(i) Assume first that the operator $\cal P$ is self-adjoint. Since, under the
above assumptions on $(M,g)$, the Laplacian $\bigtriangleup$ is self-adjoint,
this means we are treating the isovector product ${\vec A}_{i} \times
\nabla_{i}$ as a self-adjoint operator (see definition (17)), so that
$$
{\cal P}^{\dagger}=(\bigtriangleup^{\dagger}-g \Lambda^{\dagger})^{-1}
\bigtriangleup^{\dagger}
(\bigtriangleup^{\dagger}-g \Lambda^{\dagger})^{-1}={\cal P}.
\eqno (26)
$$
One can then exploit theorems ensuring that the eigenfunctions $u_{l}$ of
$\cal P$ form a complete orthonormal set. Any square-integrable function
$\varphi \in L^{2}(M)$ can be then expanded according to
($c_{l}$ being the Fourier coefficients $c_{l} \equiv (u_{l},\varphi)$)
$$
\varphi=\sum_{l=1}^{\infty}c_{l}u_{l}.
\eqno (27)
$$
The resulting mean value of $\cal P$ reads ($\lambda_{l}$ being its
eigenvalues, for which ${\cal P}u_{l}=\lambda_{l}u_{l}$)
$$
(\varphi,{\cal P}\varphi)=\sum_{l,r=1}^{\infty}
\lambda_{l}c_{r}^{*}c_{l} 
(u_{r},u_{l})=\sum_{l=1}^{\infty}\lambda_{l}|c_{l}|^{2},
\eqno (28)
$$
while more general matrix elements read (with $b_{l} \equiv 
(u_{l},\Phi)$)
$$
(\Phi,{\cal P}\varphi)=\sum_{l=1}^{\infty}\lambda_{l}b_{l}^{*}c_{l},
\eqno (29)
$$
for all $\Phi$ and $\varphi \in L^{2}(M)$. The mean value of $\cal P$ is
therefore positive if its spectrum is bounded from below, with positive
lower bound.
\vskip 0.3cm
\noindent
(ii) Even when $(M,g)$ is compact, the operator 
$\bigtriangleup -g \Lambda$ may fail to be self-adjoint, since 
$\Lambda$ contains the effect of first-order covariant derivatives, which
may be anti-self-adjoint. If this were the case, we can nevertheless exploit
the expansion (20) to point out that $\cal P$ has the general structure
$$
{\cal P}=\bigtriangleup^{-1}+{\cal P}',
\eqno (30)
$$
where ${\cal P}'$ is a pseudo-differential operator 
(see Appendix) of order $-3$, since the
lowest value of $k$ in the infinite sum in Eq. (20) involves the operator
$\bigtriangleup^{-1} \; \Lambda \; \bigtriangleup^{-1}$. One can then rely upon
the work of Beals,$^{12}$ who has studied spectral asymptotics of elliptic
operators with a self-adjoint positive principal part, showing that the
eigenvalues lie in a parabolic region around the positive axis. In our case
the principal part of $\cal P$ is the inverse of the Laplacian, which is
both self-adjoint and positive. Such a qualitative information should be
supplemented by well known results about spectral resolutions of partial
differential operators of positive order on compact manifolds.$^{13}$
In other words, lack of self-adjointness of $\cal P$ makes it now impossible
to expand $\varphi \in L^{2}(M)$ according to Eq. (27), while an expansion
of $\varphi$ remains legitimate if a discrete spectral resolution of the
Laplacian is used. For this purpose, let us denote by $v_{l}$ the eigenfunctions
of the Laplacian with discrete eigenvalues $\mu_{l}$, for which 
$$
\bigtriangleup v_{l}=\mu_{l} \; v_{l},
\eqno (31)
$$
and ($C_{l}$ being the Fourier coefficient $(v_{l},\varphi)$)
$$
\varphi=\sum_{l=1}^{\infty}C_{l}v_{l}.
\eqno (32)
$$
With the help of the expansion (20), which is useful at small $g$, the mean
value of the operator $\cal P$ is then found to take the form 
$$ \eqalignno{
(\varphi,{\cal P}\varphi)&= \sum_{l=1}^{\infty}{1\over \mu_{l}}
|C_{l}|^{2} \cr
&+ \sum_{l,r=1}^{\infty}C_{r}^{*}C_{l}\sum_{k=1}^{\infty}(k+1)g^{k}
{1\over \mu_{r}}\left(v_{r},(\Lambda \bigtriangleup^{-1})^{k}v_{l}
\right).
&(33)\cr}
$$
Such a formula shows how the mean value of $\cal P$ changes on passing from
the self-adjoint (see Eq. (28)) to the non-self-adjoint case. The formula
(33) cannot be further simplified, because the operator $\Lambda$ does
not commute with the inverse of the Laplacian. The first few powers of 
$\Lambda \bigtriangleup^{-1}$ in the sum over all positive values of
the integer $k$ yield, for example,
$$
\Lambda \bigtriangleup^{-1}v_{l}={1\over \mu_{l}}
\Lambda v_{l},
$$
$$
(\Lambda \bigtriangleup^{-1})^{2}v_{l}={1\over \mu_{l}}
\Lambda \bigtriangleup^{-1}\Lambda v_{l},
$$
and such formulae are useful if one needs to stop the analysis
of $(\varphi, {\cal P} \varphi)$ at terms of order ${\rm O}(g^{2})$. 
If the formula (33) is used, positivity of the mean value of $\cal P$
amounts to proving that the following inequality holds:
$$ 
\sum_{l=1}^{\infty}{1\over \mu_{l}}|C_{l}|^{2} >
-\sum_{l,r=1}^{\infty}C_{r}^{*}C_{l}\sum_{k=1}^{\infty}g^{k}
{1\over \mu_{r}}(v_{r},(\Lambda \bigtriangleup^{-1})^{k}v_{l}).
\eqno (34)
$$
Similarly, for all $\Phi$ and $\varphi \in L^{2}(M)$, the formula
(29) for the matrix element $(\Phi,{\cal P}\varphi)$ is now replaced
by (with our notation, $B_{l} \equiv (v_{l},\Phi)$)
$$ \eqalignno{
(\Phi,{\cal P}\varphi)&= \sum_{l=1}^{\infty}{1\over \mu_{l}}
B_{l}^{*}C_{l} \cr
&+ \sum_{l,r=1}^{\infty}B_{r}^{*}C_{l}\sum_{k=1}^{\infty}(k+1)g^{k}
{1\over \mu_{r}}(v_{r},(\Lambda \bigtriangleup^{-1})^{k}v_{l}).
&(35)\cr}
$$

The effort of studying both self-adjoint and non-self-adjoint case for
$\cal P$ is especially valuable if one looks at sub-regions of $M$ with 
smooth boundary, since then the particular boundary conditions chosen
might or might not ensure self-adjointness of $\cal P$.
\vskip 0.3cm
\leftline {\bf 4. Status of the Spectral Approach}
\vskip 0.3cm
\noindent
In our paper we start from a physical motivation to move gradually 
towards the framework of pseudo-differential operators and their relevance
for the Hamiltonian of classical Yang--Mills theory but on a background 
manifold with positive-definite metric. We have proposed
a spectral-oriented perspective on the classical foundations of an
outstanding open problem in modern quantum field theory, and we have
evaluated the well-defined matrix elements of the fundamental operator
$\cal P$ in the self-adjoint and non-self-adjoint case. This step is
important, since otherwise no hope exists of being able to evaluate the
matrix elements occurring in the quantum theory. In particular, the use 
of discrete spectral resolutions of the Laplacian for studying matrix
elements of $\cal P$ in the non-self-adjoint case is a simple but non-trivial
technical point that we have advocated.
At least three major problems are now in sight.
\vskip 0.3cm
\noindent
(i) What happens if the compact Riemannian four-manifold $M$ is
replaced by a Lorentzian four-manifold, e.g. Minkowski space-time.
Such a ``decompactification limit'' is crucial to recover the 
analysis of the original problem we are interested in. 
\vskip 0.3cm
\noindent
(ii) What happens when the classical Hamiltonian (15) is
replaced by the quantum Hamiltonian (24), with $|l,r \rangle$ being
the eigenfunctionals of the position operator.
\vskip 0.3cm
\noindent
(iii) How to make manifest the role played by the gauge group in the
quantum theory (in particular, the group structure of the operators
$\Lambda$ and $\cal P$).
\vskip 0.3cm
We hope, however, that the spectral approach will
at least show clearly the limits of what can be done to solve the
mass gap problem,$^{4,14-17}$ and it is our aim to
devote our efforts to understand whether this is really the case.
\vskip 0.3cm
\leftline {\bf Appendix}
\vskip 0.3cm
\noindent
To help the reader who is not familiar with spectral theory, we here
recall some basic concepts. A linear partial differential operator $P$
of order $d$ can be written in the form
$$
P \equiv \sum_{|\alpha| \leq d}a_{\alpha}(x)D_{x}^{\alpha},
\eqno (A1)
$$
where $|\alpha| \equiv \sum_{k=1}^{m}\alpha_{k}$, and
$$
D_{x}^{\alpha} \equiv (-i)^{|\alpha|}
{\left({\partial \over \partial x_{1}}\right)}^{\alpha_{1}} ...
{\left({\partial \over \partial x_{m}}\right)}^{\alpha_{m}},
\eqno (A2)
$$
with $a_{\alpha}$ a $C^{\infty}$ function on ${\bf R}^{m}$ for all
$\alpha$. The associated {\it symbol} is, by definition,
$$
p(x,\xi) \equiv \sum_{|\alpha| \leq d}a_{\alpha}(x)\xi^{\alpha},
\eqno (A3)
$$
i.e. it is obtained by replacing the differential operator $D_{x}^{\alpha}$
by the monomial $\xi^{\alpha}$. The pair $(x,\xi)$ may be viewed as defining
a point of the cotangent bundle of ${\bf R}^{m}$, and the action of $P$ on
the elements of the Schwarz space $\cal S$ of smooth complex-valued functions
on ${\bf R}^{m}$ of rapid decrease is given by$^{13}$
$$
Pf(x) \equiv \int e^{i(x-y)\cdot \xi}p(x,\xi)f(y)dy d\xi ,
\eqno (A4)
$$
where the $dy=dy_{1}...dy_{m}$ and $d\xi=d\xi_{1}...d\xi_{m}$ orders
of integration cannot be interchanged, since the integral is not
absolutely convergent. 

Pseudo-differential operators are instead a more general class of
operators whose symbol need not be a polynomial (and whose order is
not necessarily positive) but has suitable regularity properties.
More precisely, let $S^{d}$ be the set of all symbols $p(x,\xi)$ such
that $p$ is smooth in the pair of variables $(x,\xi)$ with compact
$x$ support, and for all $(\alpha,\beta)$ there exist constants 
$C_{\alpha,\beta}$ for which
$$
| D_{x}^{\alpha}D_{\xi}^{\beta}p(x,\xi)| \leq C_{\alpha,\beta}
(1+|\xi|)^{d-|\beta|},
\eqno (A5)
$$
for some real (not necessarily positive) value of $d$. The associated
pseudo-differential operator, defined on the Schwarz space and taking 
values in the set of smooth functions on ${\bf R}^{m}$ with compact
support,
$$
P: {\cal S} \rightarrow C_{c}^{\infty}({\bf R}^{m})
$$
is defined in a way formally analogous to the previous integral formula
for $Pf(x)$.

Ellipticity can be defined by means of a majorization obeyed by the
inverse of the modulus of the symbol. In other words, let $U$ be an
open subset with compact closure in ${\bf R}^{m}$, and consider an open
subset $U_{1}$ whose closure ${\overline U}_{1}$ is properly included in
$U: {\overline U}_{1} \subset U$. If $p$ is a symbol of order $d$ on $U$,
it is said to be elliptic on $U_{1}$ if there exists an open set $U_{2}$
which contains ${\overline U}_{1}$ and positive constants $F_{i}$ so that
$$
|p(x,\xi)|^{-1} \leq F_{1}(1+|\xi|)^{-d}
\eqno (A6)
$$
for $|\xi| \geq F_{0}$ and $x \in U_{2}$, where
$$
|\xi| \equiv \sqrt{g^{ab}(x)\xi_{a}\xi_{b}}
=\sqrt{\sum_{k=1}^{m}\xi_{k}^{2}}.
\eqno (A7)
$$
The corresponding operator $P$ is then said to be elliptic.
\vskip 0.3cm
\leftline {\bf Acknowledgements}
\vskip 0.3cm
\noindent
The work of the author has been partially
supported by the Progetto di Ricerca di Interesse Nazionale
{\sl SINTESI 2000}. He is indebted to Gerd Grubb for correspondence
on spectral theory. The INFN
financial support is also gratefully acknowledged, as well
as encouragement from Peter Orland and Pietro Santorelli. 
\vskip 0.3cm
\leftline {\bf References}
\vskip 0.3cm
\noindent
\item {[1]}
L. O'Raifeartaigh, {\it Group Structure of Gauge Theories}
(Cambridge University Press, Cambridge, 1986).
\item {[2]}
S. L. Glashow, {\it Rev. Mod. Phys.} {\bf 52} 539 (1980);  
S. Weinberg, {\it Rev. Mod. Phys.} {\bf 52}, 515 (1980);
A. Salam, {\it Rev. Mod. Phys.} {\bf 52}, 525 (1980). 
\item {[3]}
G. 't Hooft, {\it Rev. Mod. Phys.} {\bf 72}, 333 (2000); M. J. G. Veltman,
{\it Rev. Mod. Phys.} {\bf 72}, 341 (2000). 
\item {[4]}
A. M. Jaffe and E. Witten, ``Quantum Yang--Mills Theory'', Clay Mathematics
Institute Millennium Prize Problem, 2000.
\item {[5]}
F. E. Close and P. R. Page, {\it Sci. Am.} {\bf 279}, 52 (1998). 
\item {[6]}
F. E. Close, {\it Acta Phys. Polon.} {\bf B31}, 2557 (2000). 
\item {[7]}
T. D. Lee, {\it Particle Physics and Introduction to Field Theory}
(Harwood Academic Publishers, New York, 1980); N. H. Christ and
T. D. Lee, {\it Phys. Rev.} {\bf D22}, 939 (1980). 
\item {[8]}
L. Baulieu and D. Zwanziger, {\it Nucl. Phys.} {\bf B548}, 527 (1999). 
\item {[9]}
A. S. Wightman, {\it Fortschr. Phys.} {\bf 44}, 143 (1996).
\item {[10]}
J. Glimm and A. Jaffe, {\it Quantum Physics. A Functional Integral
Point of View} (Springer--Verlag, New York, 1987).
\item {[11]}
G. Esposito, A. Yu. Kamenshchik and G. Pollifrone, {\it Euclidean
Quantum Gravity on Manifolds with Boundary} (Kluwer, Dordrecht, 1997).
\item {[12]}
R. Beals, {\it Am. J. Math.} {\bf 89}, 1056 (1967).
\item {[13]}
P. B. Gilkey, {\it Invariance Theory, the Heat Equation and the
Atiyah--Singer Index Theorem} (CRC Press, Boca Raton, 1995).
\item {[14]}
I. M. Singer, {\it Physica Scripta} {\bf 24}, 817 (1980).
\item {[15]}
O. Babelon and C. M. Viallet, {\it Commun. Math. Phys.} {\bf 81},
515 (1981).
\item {[16]}
R. P. Feynman, {\it Nucl. Phys.} {\bf B188}, 479 (1981). 
\item {[17]}
P. Orland, The Metric on the Space of Yang--Mills Configurations
(HEP-TH 9607134).
\end{thebibliography}

\bye